\documentclass[preprint,authoryear]{elsarticle_mod}

\usepackage[latin1]{inputenc}
\usepackage[english]{babel} 
\usepackage{subfigure}
\usepackage{amsmath,amsfonts}
\usepackage{setspace}
\usepackage{tabularx} 
\usepackage{comment}
\usepackage{color}
\usepackage{url}

\allowdisplaybreaks[4] 
\newcommand{\R}{\textit{I\!R}}
\newcommand{\tablesize}{\fontsize{7.5}{10}\selectfont}

\usepackage{amssymb}
\usepackage{amsthm}

\begin{document}

\begin{frontmatter}



\title{Bootstrap-based model selection criteria for beta regressions}


 
  \author[label1]{F\'abio~M.~Bayer\corref{cor1}}
 \ead{bayer@ufsm.br}
 \address[label1]{Departamento de Estatística and LACESM, Universidade Federal de Santa Maria}
 \cortext[cor1]{Corresponding author}  
 
 \author[label2]{Francisco~Cribari-Neto}
 \ead{cribari@de.ufpe.br}
 \address[label2]{Departamento de Estatística, Universidade Federal de Pernambuco}



\onehalfspacing

\begin{abstract}
The Akaike information criterion (AIC) is a model selection criterion 
widely used in practical applications. The AIC is an estimator of the log-likelihood expected value, and measures the discrepancy between the true model and the estimated model. In small samples the AIC is biased and tends to select overparameterized models. To circumvent that problem, we propose two new selection criteria, namely: the bootstrapped likelihood quasi-CV (BQCV) and its 632QCV variant. 
We use Monte Carlo simulation to compare the finite sample performances of the two proposed criteria to those of the AIC and its variations that use the bootstrapped log-likelihood in the class of varying dispersion beta regressions. 
The numerical evidence shows that the proposed model selection criteria perform well in small samples. We also present and discuss and empirical application.

\end{abstract}


\end{frontmatter}


\onehalfspacing

\section{Introduction}

In regression analysis, practitioners are usually interested in selecting the model that yields the best fit from a broad class of candidate models. Thus, model selection is of paramount importance in regression analysis. Model selection is usually based on model selection criteria or information criteria. The Akaike information criterion (AIC) \citep{Akaike1973} is the most well-known and commonly used model selection criterion. Several alternative criteria have been developed in the literature, such as the SIC \citep{Schwarz1978}, HQ \citep{Hannan1979} e AICc \citep{Hurvich1989}.

The AIC was proposed for estimating (minus two times) the expected log-likelihood. Using Taylor series expansion and the asymptotic normality of the maximum likelihood estimator Akaike showed that the maximized log-likelihood function is a positively biased estimator for the expected log-likelihood. After computing such bias the author derived the AIC as an asymptotically approximated correction for the expected log-likelihood. In small samples, however, the AIC is biased and tends to select models that are overparameterized \citep{Hurvich1989}.

Several variants of the AIC have been proposed in the literature. The first correction of the AIC, the AICc, was proposed in \cite{Sugiura1978} for linear regression models. Later, \cite{Hurvich1989} expanded the applicability of the AICc to cover nonlinear regression and autoregressive models. They showed that the AICc is asymptotically equivalent to the AIC but usually delivers more accurate model selection in finite samples. Analytical corrections to the AIC, such as AICc, can be nonetheless difficult to obtain in some classes of models  \citep{Shibata1997}. The analytical difficulties stem from distributional and asymptotic results, as well as from certain restrictive assumptions. In order to circumvent analytical difficulties and to obtain more accurate corrections in small samples, bootstrap \citep{Efron1979} variants of the AIC were considered in the literature. They have been introduced and explored in different classes of models. See, for instance, \cite{Cavanaugh1997}, \cite{Ishiguro1991}, \cite{Ishiguro1997}, \cite{Seghouane2010}, \cite{Shang2008} and \cite{Shibata1997}, who introduced the criteria known as WIC, AICb, EIC, among other denominations. Such bootstrap extensions typically outperform the AIC in finite samples. Additionally, as noted by \cite{Shibata1997}, they can be easily computed. 

Both the AIC and its bootstrap variants aim at estimating the expected log-likelihood using a bias correction for the maximized log-likelihood. In this paper, we follow the approach introduced by \cite{Pan1999} and propose an estimator for the expected log-likelihood that does not require a bias adjustment term. In particular, nonparametric bootstrap and cross validation (CV) are jointly used in a criterion called bootstrapped likelihood CV (BCV). Using the parametric bootstrap and a quasi-CV method  we define a new AIC variant. It uses the bootstrapped likelihood quasi-CV (BQCV). We also propose a slice modification known as 632QCV. 

Model selection criteria based on the bootstrapped log-likelihood have been explored and successfully applied to autoregressive models \citep{Ishiguro1997}, state-space models \citep{Bengtsson2006, Cavanaugh1997}, mixed models \citep{Shang2008}, linear regression models \citep{Pan1999, Seghouane2010} and logistic and Cox regression models \citep{Pan1999}. In this paper, we investigate model selection via bootstrap log-likelihood in the class of beta regression models. Such models were introduced by \cite{Ferrari2004} and are tailored for modeling responses that assume values in the standard unit interval, $(0,1)$, such as rates and proportions. We consider the class of varying dispersion beta regressions, as described in \cite{Simas2010}, \cite{Pinheiro2011} and \cite{CribariSouza2012}. It generalizes the fixed dispersion beta regression model proposed by \cite{Ferrari2004}. The model has two submodels, one for the mean and another one for the dispersion. 

The chief goal of our paper is twofold. First, we propose new model selection criteria for beta regressions and then we numerically investigate their finite sample performances in small samples. We also provide simulation results on alternative model selection strategies. The numerical evidence shows that the criteria we propose typically yield reliable model selection in the class of beta regression models.  

This paper is organized as follows. In the next section we introduce the AIC and its bootstrap extensions. We also propose two new model selection criteria. Section~\ref{S:modelo_reg_beta} introduces the class of beta regression models. In Section~\ref{S:avaliacao_AIC} we present Monte Carlo simulation results on model selection in fixed and varying beta regression models. An empirical application is presented and discussed in Section~\ref{S:application_BQCV}. Finally, some concluding remarks are offered in Section~\ref{S:conclusions}.

\section{Akaike information criterion and bootstrap variations}

The distance measure between two densities can be measured using the Kullback-Leibler (KL) information \citep{Kullback1968}, also known as entropy or discrepancy \citep{Cavanaugh1997b}. The KL information can be used to select an estimated model which is closest to the true model. The AIC was derived by \cite{Akaike1973} by minimizing the KL information. In what follows, we shall follow \cite{Bengtsson2006} in order to formalize the notion of selecting a model from a class of candidate models.  

Suppose the $n$-dimensional vector $Y$ is sampled from an unknown density $f(Y|\theta_{k_0})$, where $\theta_{k_0}$ is a $k_0$-vector of parameters. The respective parametric family of densities is denoted by $\mathcal{F}(k_i)=\left\{f(Y|\theta_{k_i})|\theta_{k_i}\in\Theta_{k_i}\right\}$, where $\Theta_{k_i}$ is the $k_i$-dimensional parametric space. Let $\hat{\theta}_{k_i}$ be the maximum likelihood estimate of $\theta_{k_i}$. It is obtained by maximizing $f(Y|\theta_{k_i})$ in $\Theta_{k_i}$, i.e., $f(Y|\hat{\theta}_{k_i})$ is the maximized likelihood function.

Using the AIC it is possible to select the model that best approximates $f(Y|\theta_{k_0})$ from the class of families $\mathcal{F}=\{\mathcal{F}(k_1), \mathcal{F}(k_2), \ldots $, $ \mathcal{F}(k_L)\}$, $i=1,2,\ldots,L$. For notation simplicity we will not consider different families in the class $\mathcal{F}$ which have the same dimension. We say that $f(Y|\hat{\theta}_k)$ is correctly specified if $f(Y|\theta_{k_0})\in\ \mathcal{F}(k)$, where $\mathcal{F}(k)$ is the smallest dimensional family that contains $f(Y|\theta_{k_0})$. We say that $f(Y|\hat{\theta}_k)$ is overspecified if $f(Y|\theta_{k_0})\in \mathcal{F}(k)$ but families of smaller dimension also contain $f(Y|\theta_{k_0})$. On the other hand, $f(Y|\hat{\theta}_k)$ is underspecified if $f(Y|\theta_{k_0})\notin \mathcal{F}(k)$.

The KL measure can be used to determine which fitted model (i.e.,  which model in the collection $f(Y|\hat{\theta}_{k_1}), f(Y|\hat{\theta}_{k_2}), \ldots, f(Y|\hat{\theta}_{k_L})$) is closest to  $f(Y|\theta_{k_0})$. The KL distance between the true model $f(Y|\theta_{k_0})$ and the candidate model $f(Y|\theta_k)$ is given by
\begin{equation*}
d(\theta_{k_0},\theta_k) = {\rm E}_0\!\left[\log\left\{\frac{f(Y|\theta_{k_0})}{f(Y|\theta_{k})} \right\} \right],
\end{equation*}
where ${\rm E}_0(\cdot)$ denotes expectation under $f(Y|\theta_{k_0})$. Let
\begin{equation}\label{E:delta}
\delta(\theta_{k_0},\theta_k) = {\rm E}_0\!\left\{-2\log f(Y|\theta_{k}) \right\}.
\end{equation}
It is possible to show that $2d(\theta_{k_0},\theta_k) = \delta(\theta_{k_0},\theta_k) - \delta(\theta_{k_0},\theta_{k_0})$. Since $\delta(\theta_{k_0},\theta_{k_0})$ does not depend on $\theta_k$ minimizing $2d(\theta_{k_0},\theta_k)$ or $d(\theta_{k_0},\theta_k)$ is equivalent to minimizing the discrepancy $\delta(\theta_{k_0},\theta_k)$. Therefore, the model $f(Y|\theta_{k})$ that minimizes minus two times the expected log-likelihood, $\delta(\theta_{k_0},\theta_k)$, is the closest model to the true model according to the Kullback-Leibler information.

Notice that
\begin{equation*}
\delta(\theta_{k_0},\hat{\theta}_k) = {\rm E}_0\!\left\{-2\log f(Y|\theta_{k}) \right\}|_{\theta_{k}=\hat{\theta}_k}
\end{equation*}
measures the distance between the true model and the estimated candidate model. However, it is not possible to evaluate $\delta(\theta_{k_0},\hat{\theta}_k)$, since it requires knowledge of density $f(Y|\theta_{k_0})$. \cite{Akaike1973} used $-2\log f(Y|\hat{\theta}_k)$ as an estimator for $\delta(\theta_{k_0},\hat{\theta}_k)$. Its bias 
\begin{equation}\label{E:vies}
B={\rm E}_0\left\{-2\log f(Y|\hat{\theta}_k)-\delta(\theta_{k_0},\hat{\theta}_k)\right\}
\end{equation}
can be asymptotically approximated by $-2k$, where $k$ is the dimension of $\theta_k$.

Thus, the expected value of Akaike's criterion, 
\begin{equation*}
{\rm AIC} = -2\log f(Y|\hat{\theta}_k)+2k,
\end{equation*}
is asymptotically equal to the expected value of $\delta(\theta_{k_0},\hat{\theta}_k)$, which is given by 
\begin{equation*}
\Delta(\theta_{k_0},k)= {\rm E}_0\left\{\delta(\theta_{k_0},\hat{\theta}_k)\right\}.
\end{equation*}
Notice that $-2\log f(Y|\hat{\theta}_k)$ is a biased estimator of minus two times the expected log-likelihood and the penalizing term of the AIC, $2k$, is an adjustment term for the bias given in (\ref{E:vies}).

Since the AIC is based on a large sample approximation it may perform poorly in small samples \citep{Bengtsson2006}. Several variants of the AIC were developed aiming at delivering more accurate model selection in small samples.  \cite{Sugiura1978} developed the AICc, which in class of linear regression models is an unbiased estimator of $\Delta(\theta_{k_0},k)$, that is, ${\rm E}_0\left\{{\rm AICc}\right\}=\Delta(\theta_{k_0},k)$. Based on the results obtained by \cite{Sugiura1978}, \cite{Hurvich1989} extended the use of the AICc to cover nonlinear regression and for autoregressive models. The authors showed that the AICc is asymptotically equivalent to the AIC, i.e., ${\rm E}_0\left({\rm AICc}\right)+ o(1)=\Delta(\theta_{k_0},k)$, and typically outperforms the AIC in small samples.

According to \cite{Cavanaugh1997b}, the advantage of AICc over the AIC is that the former estimates the expected discrepancy more accurately than the latter. On the other hand, a clear advantage of the AIC over the AICc is that the AIC is universally applicable, regardless of the class of models, whereas the AICc derivation is model dependent.

\subsection{Bootstrap extensions of AIC}

Bootstrap extensions of AIC (EIC) are criteria that use bootstrap estimators for the bias term $B$ given in (\ref{E:vies}). They typically include a bias estimate which is more accurate than $-2k$ in small samples, thus leading to more reliable model selection. In what follows, we shall use five different bootstrap estimators, $B_i$ ($i=1,\ldots,5$) for $B$. The bias estimator $B_i$ defines five bootstrap extensions of AIC which we denote by ${\rm EIC}i$, $i=1,\ldots,5$. The bootstrap variants of the AIC that we shall use for model selection in the class of beta regressions have the following form:
\begin{equation*}
{\rm EIC}i= -2\log f(Y|\hat{\theta}_k) + B_i, \quad i=1,\ldots,5.
\end{equation*}

 Let $Y^*$ be a bootstrap sample (generated either parametrically or nonparametrically) and let ${\rm E}_*$ denote the expected value with respect to distribution of $Y^*$. Consider $W$ bootstrap samples $Y^*(i)$ and the corresponding estimates of $\hat{\theta}_k$: $\left\{\hat{\theta}^*_k(i)\right\}$, $i=1,\,2,\,\ldots,\,W$. Here, each estimate $\hat{\theta}^*_k(i)$ is the value of $\theta_k$ that maximizes the likelihood function $f(Y^*(i)|\theta_k)$. 
 
\cite{Ishiguro1997} proposed a bootstrap extension of the AIC known as the EIC. It is a particular case of the WIC \citep{Ishiguro1991} obtained considering independent and identically distributed (i.i.d.) observations. We shall refer to such a criterion as ${\rm EIC1}$. It estimates the bias in  (\ref{E:vies}) as
\begin{equation*}
B_1={\rm E}_*\left\{2\log f(Y^*|\hat{\theta}_k^*)-2\log f(Y|\hat{\theta}_k^*)\right\}.
\end{equation*}

A different bootstrap-based criterion was proposed in \cite{Cavanaugh1997} for the selection of state-space models; we shall refer to it as ${\rm EIC2}$. The criterion estimates the bias in (\ref{E:vies}) as
\begin{equation*}
B_2 = 2{\rm E}_*\left\{2\log f(Y|\hat{\theta}_k)-2\log f(Y|\hat{\theta}_k^*)\right\}.
\end{equation*}

We note that ${\rm EIC1}$ and ${\rm EIC2}$ are called AICb1  and AICb2, respectively, in \cite{Shang2008} in the context of mixed models selection based on the parametric bootstrap.

\cite{Shibata1997} showed that $B_1$ and $B_2$ are asymptotically equivalent and proposed the following three bootstrap estimators of (\ref{E:vies}):
\begin{equation*}
B_3=2{\rm E}_*\left\{2\log f(Y^*|\hat{\theta}_k^*)-2\log f(Y^*|\hat{\theta}_k)\right\},
\end{equation*}
\begin{equation*}
B_4=2{\rm E}_*\left\{2\log f(Y^*|\hat{\theta}_k)-2\log f(Y|\hat{\theta}_k^*)\right\},
\end{equation*}
\begin{equation*}
B_5=2{\rm E}_*\left\{2\log f(Y^*|\hat{\theta}_k^*)-2\log f(Y|\hat{\theta}_k)\right\}.
\end{equation*}
We shall refer to the corresponding criteria as ${\rm EIC3}$, ${\rm EIC4}$ and ${\rm EIC5}$. 

\cite{Seghouane2010} proposed corrected versions of the AIC for the linear regression model as asymptotic approximations to ${\rm EIC1}$, ${\rm EIC2}$, ${\rm EIC3}$, ${\rm EIC4}$ and ${\rm EIC5}$ obtained using the parametric bootstrap.

\subsection{Bootstrapped likelihood and cross-validation}

The model selection criteria described so far aim at estimating the expected log-likelihood using a bias correction for the maximized log-likelihood function.  \cite{Pan1999}, however, tried to obtain an estimator for the expected log-likelihood that does not require a bias adjustment. It uses cross-validation (CV) and bootstrap.

 CV is widely used for estimating the error rate of prediction models \citep{Efron1983, Efron1997}. In the context of model selection, according to \cite{Davies2005}, the first CV based criterion was the PRESS \citep{Allen1974}. Bootstrap based model selection was introduced by \cite{Efron1986}.  \cite{Breiman1992} and \cite{Hjorth1994} discuss the use of CV and bootstrap in model selection.
 
According to \cite{Efron1983} and \cite{Efron1997}, CV typically reduces bias but leads to variance inflation. Such variability can be reduced by using the bootstrap method. In the context of model selection of models, \cite{Pan1999} introduced a method that combines nonparametric bootstrap and CV: the bootstrapped likelihood CV (BCV). BCV yields an estimator of (\ref{E:delta}) that does not entail bias correction. For a sample $Y$ of size $n$, the BCV is defined by
\begin{equation*}
{\rm BCV} = {\rm E}_*\left\{-2\log f(Y^-|\hat{\theta}_k^*)\frac{n}{m^*}\right\},
\end{equation*}
where $Y^*$ is the bootstrap sample generated nonparametrically, $Y^-\! = Y\!-\!Y^*$, that is, $Y=Y^- \!\cup Y^*$ and $Y^- \! \cap Y^* \!= \emptyset $, and $m^*\!>0$\ is the number of elements of $Y^-$. Thus, no observation of $Y$ is used twice: each observation either belongs to $Y^*$ or to $Y^-$.

Following \cite{Efron1983}, \citeauthor{Pan1999} argues that the BCV can overestimate (\ref{E:delta}) and, on the other hand, $-2\log f(Y|\hat{\theta}_k)$ may underestimate it. Thus, following the 632+ rule \citep{Efron1997}, \citeauthor{Pan1999} introduces the 632CV criterion as
\begin{equation*}
{\rm 632CV} = 0.368\left\{-2\log f(Y|\hat{\theta}_k) \right\} + 0.632 {\rm BCV}.
\end{equation*}

\subsection{Proposed bootstrapped likelihood quasi-CV}

We shall now introduce two new model selection criteria of models that incorporate corrections for small samples. Like the BCV, these criteria provide direct estimators for the expected log-likelihood.

 Let $F$ be the distribution function of the observed sample $Y=(y_1,\dots,y_n)$ and let $\hat{F}$ be the estimated distribution function, i.e., $\hat{F}$ is the distribution function $F$ evaluated at the estimative $\hat{\theta}$. We define
\begin{eqnarray}
Y^*_p = (y^*_1,y^*_2,\dots,y^*_n) & \!\!\!\! \sim & \!\!\!\! \hat{F} \quad \text{estimation sample (or training sample),}\\
Y=(y_1,y_2,\dots,y_n )& \!\!\!\! \sim & \!\!\!\! F \quad \text{validation sample.}
\end{eqnarray}
Suppose we have $W$ pseudo-samples $Y^*_p$ obtained from $\hat{F}$ and let $\{\hat{\theta}^{p*}_k(i),\, i=1,2,\ldots,W\}$ denote the set of $W$ bootstrap replications of $\hat{\theta}_k$. We define the bootstrapped likelihood quasi-CV (BQCV) criterion as follows:
\begin{equation*}
{\rm BQCV} = {\rm E}_{p*}\left\{-2\log f(Y|\hat{\theta}_k^{p*})\right\},
\end{equation*}
where ${\rm E}_{p*}$ is the expected value with respect to the distribution of $Y^*_p$.

It follows from the strong law of large numbers that
\begin{equation*}
\frac{1}{W}\sum_{i=1}^{W}\left\{-2\log f(Y|\hat{\theta}_k^{p\ast}(i))\right\} \xrightarrow[W \rightarrow \infty]{a.s.} {\rm E}_{p\ast}\left\{-2\log f(Y|\hat{\theta}_k^{p\ast})\right\},\label{E:as1}
\end{equation*}
where $\xrightarrow{a.s.}$ denotes almost sure convergence.

The computation of BQCV can be performed as follows: 
\begin{enumerate}
\item Estimate $\theta$ using the sample $Y=(y_1,\ldots,y_n)$;
\item Generate $W$ pseudo-samples $Y^*_p$ from $\hat{F}$;
\item For each $Y^*_p(i)$, $i=1,\ldots,W$, compute $\hat{\theta}^{p*}_k(i)$ and $-2\log f(Y|\hat{\theta}_k^{p\ast}(i))$;
\item Using the $W$ replications of $-2\log f(Y|\hat{\theta}_k^{p\ast})$ compute  
\begin{equation*}
{\rm BQCV} = \frac{1}{W}\sum_{i=1}^{W}\left\{-2\log f(Y|\hat{\theta}_k^{p\ast}(i))\right\}.
\end{equation*}
\end{enumerate}
Based on pilot simulations, we recommend using $W=200$. 

The algorithm outlined above is not a genuine cross-validation scheme, hence the name quasi-CV. It is not a genuine cross-validation scheme because it does not partition the sample $Y$, but instead it treats the samples $Y^*_p$ and $Y$ as partitions of the same data set. In each bootstrap replication we use a procedure which is similar to the twofold CV. Here, the training sample is the pseudo-sample of the parametric bootstrap scheme, $Y^*_p$, and the validation sample is the observed sample, $Y$. 

Following the approach used by \cite{Pan1999} for obtaining the 632CV, we propose another model selection criterion, which we call 632QCV. It is a variant of the BQCV and is given by 
\begin{equation*}
{\rm 632QCV} = 0.368\left\{-2\log f(Y|\hat{\theta}_k) \right\} + 0.632 {\rm BQCV}.
\end{equation*}

\section{The beta regression model}\label{S:modelo_reg_beta}

Many studies in different fields examine how a set of covariates is related to a response variable that assumes values in continuous interval, $(0,1)$, such as rates and proportions; see, e.g., \cite{Brehm1993}, \cite{Hancox2010}, \cite{Kieschnick2003}, \cite{Ferrari2004}, \cite{Smithson2006}, \cite{Zucco2008}, \cite{Verhaelen2013}, \cite{Whiteman2013} and \cite{Hallgren2013}. Such modeling can be done using the class of beta regression models, which was introduced by \cite{Ferrari2004}. It assumes that the response variable ($y$) follows the beta law. The beta distribution is quite flexible since its density can assume a number of different shapes depending on the parameter values. The beta density can be indexed by mean ($\mu$) and dispersion ($\sigma$) parameters when written as  
\begin{equation}\label{E:densidade2}
f(y|\mu,\sigma)\!=\!
\frac{\Gamma\left(\frac{1-\sigma^2}{\sigma^2}\right)}{\Gamma\left(\mu\left(\frac{1-\sigma^2}{\sigma^2}\right)\right)\Gamma\left(\left(1-\mu\right)\left(\frac{1-\sigma^2}{\sigma^2}\right)\right)}y^{\mu\left(\frac{1-\sigma^2}{\sigma^2}\right)-1}(1\!-\!y)^{(1-\mu)\left(\frac{1-\sigma^2}{\sigma^2}\right)-1}, 
\end{equation}
where $0<y<1$, $0<\mu<1$, $0<\sigma<1$, $\Gamma{(\cdot)}$ is the gamma function and $V(\mu)=\mu (1-\mu)$ is the variance function. The mean and the variance of $y$ are, respectively, by ${\rm E}(y)=\mu$ and ${\rm var}(y)=V(\mu)\sigma^2$.

Let $Y=(y_1,\ldots,y_n)$ be a vector of independent random variables, where $y_t$, $t=1,\ldots,n$, has density (\ref{E:densidade2}) with mean $\mu_t$ and unknown dispersion $\sigma_t$. The varying dispersion beta regression model can be written as
\begin{align}
& g(\mu_t)=\sum_{i=1}^{r}x_{ti}\beta_i=\eta_t,\label{E:sm_mean} \\
& h(\sigma_t)=\sum_{i=1}^{s}z_{ti}\gamma_i=\nu_t, \label{E:sm_disp}
\end{align}
where ${\beta}=(\beta_1,\ldots,\beta_r)^{\top}$ and ${\gamma}=(\gamma_1,\ldots,\gamma_s)^{\top}$ are vectors of unknown parameters and  $x_t=(x_{t1},\ldots,x_{tr})^{\top}$ and $z_t=(z_{t1},\ldots,z_{ts})^{\top}$ are observations on $r$ and $s$ independent variables, $r+s=k<n$. 
In what follows, we denote the matrix of regressors used in the mean submodel by $X$, i.e., $X$ is the $n\times r$ matrix whose $t$th line is $x_t$. Likewise, $Z$ is the matrix of regressors used in the dispersion submodel. When intercepts are included in the mean and dispersion submodels, $x_{t1}=z_{t1}=1$, for $t=1,\ldots,n$. Additionally, $g(\cdot)$ and $h(\cdot)$ are strictly monotonic and twice differentiable link functions with domain in $(0,1)$ and image in \R. In the parameterization we use, the same link functions can be used in the mean and dispersion submodels. Commonly used link functions are logit, probit, log-log, complementary log-log and Cauchy. A detailed discussion of link functions can be found in \cite{McCullagh1989} and \cite{Koenker2009}. Finally, we note that the constant dispersion beta regression model is obtained by setting $s=1$, $z_{t1}=1$ and $h(\cdot)$ is the identity function.

Joint estimation of ${\beta}$ and ${\gamma}$ can be performed by maximum likelihood. 
Let $\theta_k = (\beta_1, \ldots, \beta_r, \gamma_1, \ldots, \gamma_s)^{\top}$  and let be $Y$ an $n$-vector of independent beta random variables. The log-likelihood function is 
\begin{equation*}
\log f(Y|\theta_k)=\sum^n_{t=1}\log f(y_t|\mu_t,\sigma_t),
\end{equation*}
where
\small
\begin{align*}
\log f(y_t|\mu_t,\sigma_t) & 
= \log \Gamma{\left( \!\frac{1-\sigma^2_t}{\sigma^2_t} \right)}\!\!- \log \Gamma{\left(\! \mu_t \!\left(\!\frac{1-\sigma^2_t}{\sigma^2_t}\right)\! \!\right)}\!- \log \Gamma{\left(\!\!(1\!-\mu_t)\!\left(\frac{1-\sigma^2_t}{\sigma^2_t} \right) \!\!\right)} \\
& 
+ \left[\mu_t \!\left(\frac{1-\sigma^2_t}{\sigma^2_t} \right)\!-1 \right] \log y_t + \left[(1-\mu_t)\left(\frac{1-\sigma^2_t}{\sigma^2_t}\right)\!-1 \right]\log(1-y_t).
\end{align*}
\normalsize

The score function is obtained  by differentiating the log-likelihood function with respect to the unknown parameters. Closed-form expressions for the score function and Fisher's information matrix are given in \ref{A:escore_fisher}.

Let $U_{\!\beta}({\beta},{\gamma})$  and $U_{\!\gamma}({\beta},{\gamma})$ be the score functions for ${\beta}$ and ${\gamma}$, respectively. The maximum likelihood estimators are obtained by solving 
\begin{equation*}
\left\{
 \begin{array}{ll}
  U_{\!\beta}({\beta},{\gamma})= & \! 0, \\
  U_{\!\gamma}({\beta},{\gamma})= & \! 0.
 \end{array}
\right.
\end{equation*}
The solution to such a system of equations does not have a closed form. Hence, maximum likelihood estimates are usually obtained by numerically maximizing the log-likelihood function. 

A global goodness-of-fit measure can be obtained by transforming the likelihood ratio as \citep{Nagelkerke1991}
\begin{equation*}
R^2_{LR} = 1-\left(\frac{L_{\rm null}}{L_{\rm fit}}\right)^{2/n},
\end{equation*}
where $L_{\rm null}$ is the maximized likelihood function of the model without regressors and $L_{\rm fit}$ is the maximized likelihood function of the fitted regression model. An alternative measure is the square of the correlation coefficient between $g( {y})$ and $\widehat{ {\eta}}=X\widehat{ {\beta}}$, where $\widehat{ {\beta}}$ denotes the maximum likelihood estimator of ${\beta}$. Such a measure, which we denote by $R^2_{FC}$, was proposed by \cite{Ferrari2004} for constant dispersion beta regressions.

\section{Numerical evaluation}\label{S:avaliacao_AIC}

In this section we investigate the performances of the AIC and its bootstrap variations in small samples when used in the selection of beta regression models. All simulations were performed using the Ox matrix programming language \citep{Doornik2007}. All log-likelihood maximizations were numerically carried out using the  quasi-Newton nonlinear optimization algorithm known as BFGS with analytic first derivatives.\footnote{For details on the BFGS algorithm, see \cite{Press1992}.}

We consider beta regression models with mean submodel as given in (\ref{E:sm_mean}) and dispersion submodel as given in (\ref{E:sm_disp}). 
We used $1000$ Monte Carlo replications and, for each sample, $W=200$ bootstrapped log-likelihoods were computed. We experimented with larger values of $W$ but noticed that they only yielded negligible improvements in the model selection criteria performances. For the bootstrap extensions of AIC we investigated the use of the parametric bootstrap, ${\rm EIC}i_{p}$, as well as the use of the nonparametric bootstrap, ${\rm EIC}i_{np}$. We also considered alternative model selection criteria in the Monte Carlo simulations: AICc \citep{Hurvich1989}, SIC \citep{Schwarz1978}, SICc \citep{McQuarrie1999}, HQ \citep{Hannan1979} and HQc \citep{McQuarrie1998}.\footnote{The use of these criteria in beta regression models is done in an \textit{ad hoc} manner.} The covariates values were obtained as random $\mathcal{U}(0,1)$ draws; they were kept constant throughout the experiment. The logit link function was used in both submodels.  


Performance evaluation of the different criteria is done as in \cite{Hannan1979}, \cite{Hurvich1989}, \cite{Shao1996}, \cite{McQuarrie1997}, \cite{McQuarrie1998}, \cite{Pan1999}, \cite{Shi2002}, \cite{Davies2005}, \cite{Shang2008}, \cite{Hu2008}, \cite{Liang2008}, and \cite{Seghouane2010}.
For each criterion we present the frequency of correct order selection ($=k_0$), as well as the frequencies of underspecified ($<k_0$) and overspecified ($>k_0$) selected models. 

The following data generating processes were used: 
\begin{eqnarray}
{\rm logit}(\mu_t)=-1.5 + x_{t2} + x_{t3}, &  {\rm logit}(\sigma_t)=-0.7-0.6x_{t2} -0.6 x_{t3}, \label{E:model5} \\
{\rm logit}(\mu_t)= 1 - 0.75 x_{t2} - 0.25 x_{t3}, &  {\rm logit}(\sigma_t)=-0.7-0.5x_{t2} -0.3 x_{t3}, \label{E:model6} \\
{\rm logit}(\mu_t)=-1.5 + x_{t2} + x_{t3}, &  {\rm logit}(\sigma_t)=-1.1-1.1x_{t2} -1.1 x_{t3}, \label{E:model7} \\
{\rm logit}(\mu_t)= 1 - 0.75 x_{t2} - 0.25 x_{t3}, & \! {\rm logit}(\sigma_t)=-1.45-1x_{t2} -0.5 x_{t3}. \label{E:model8}
\end{eqnarray}
The first two models, (\ref{E:model5}) and (\ref{E:model6}), entail large dispersion whereas the remaining two models, (\ref{E:model7}) and (\ref{E:model8}), have small dispersion. Considering the parameters values, we note that the regression models in (\ref{E:model5}) and (\ref{E:model7}) are easily identifiable whereas the models in (\ref{E:model6}) and (\ref{E:model8}) are weakly identifiable. 
In the weak identifiability scenario, variations in the covariates have different impacts on the mean response. 
The terminology ``easily identified models'' is used here in the same sense as in  \cite{McQuarrie1998}, \cite{Caby2000} and \cite{Frazer2009}.
We emphasize that such a concept of model identifiability differs from the usual concept which  
relates to the model parameters uniqueness \citep{Paulino1994, Rothenberg1971}.
The numerical results for models with large and small dispersion are similar and, for that reason, we only present results for models with small dispersion, (\ref{E:model7}) and (\ref{E:model8}).

\begin{table}[t]
\tablesize
\caption{Frequencies of correct and incorrect order selection from 1000 independent replications; mean and dispersion regressors jointly selected in an easily identified model (Model (\ref{E:model7})). 
} \label{T:M1_b3}
{
\begin{center}
\begin{tabularx}{\textwidth}{ l|XXX|XXX|XXX|XXX}	
\hline
 & \multicolumn{3}{c|}{$n=25$} & \multicolumn{3}{c|}{$n=30$} & \multicolumn{3}{c|}{$n=40$} & \multicolumn{3}{c}{$n=50$}\\

	& $	<k_0	$ & $	=k_0	$ & $	>k_0	$ & $	<k_0	$ & $	=k_0	$ & $	>k_0	$ & $	<k_0	$ & $	=k_0	$ & $	>k_0	$ & $	<k_0	$ & $	=k_0	$ & $	>k_0	$ \\
\hline

AIC	& $	195	$ & $	100	$ & $	705	$ & $	229	$ & $	169	$ & $	602	$ & $	181	$ & $	273	$ & $	546	$ & $	156	$ & $	345	$ & $	499	$ \\
AICc	& $	618	$ & $	167	$ & $	215	$ & $	532	$ & $	233	$ & $	235	$ & $	329	$ & $	373	$ & $	298	$ & $	242	$ & $	464	$ & $	294	$ \\
SIC	& $	476	$ & $	122	$ & $	402	$ & $	557	$ & $	190	$ & $	253	$ & $	518	$ & $	312	$ & $	170	$ & $	439	$ & $	423	$ & $	138	$ \\
SICc	& $	883	$ & $	72	$ & $	45	$ & $	864	$ & $	99	$ & $	37	$ & $	718	$ & $	237	$ & $	45	$ & $	607	$ & $	337	$ & $	56	$ \\
HQ	& $	274	$ & $	121	$ & $	605	$ & $	325	$ & $	191	$ & $	484	$ & $	288	$ & $	333	$ & $	379	$ & $	253	$ & $	436	$ & $	311	$ \\
HQc	& $	734	$ & $	128	$ & $	138	$ & $	671	$ & $	192	$ & $	137	$ & $	507	$ & $	349	$ & $	144	$ & $	386	$ & $	457	$ & $	157	$ \\
BQCV	& $	861	$ & $	107	$ & $	32	$ & $	640	$ & $	267	$ & $	93	$ & $	309	$ & $	\mathbf{466} $ & $	225	$ & $	187	$ & $	\mathbf{506}	$ & $	307	$ \\
632QCV	& $	678	$ & $	\mathbf{234}	$ & $	88	$ & $	387	$ & $	\mathbf{371}	$ & $	242	$ & $	151	$ & $	420 $ & $	429	$ & $	80	$ & $	362	$ & $	558	$ \\
$ {\rm EIC1}_p $	& $	964	$ & $	28	$ & $	8	$ & $	950	$ & $	28	$ & $	22	$ & $	893	$ & $	77	$ & $	30	$ & $	886	$ & $	73	$ & $	41	$ \\
$ {\rm EIC2}_p $	& $	980	$ & $	19	$ & $	1	$ & $	920	$ & $	63	$ & $	17	$ & $	722	$ & $	249	$ & $	29	$ & $	515	$ & $	419	$ & $	66	$ \\
$ {\rm EIC3}_p $	& $	856	$ & $	129	$ & $	15	$ & $	521	$ & $	368	$ & $	111	$ & $	267	$ & $	422	$ & $	311	$ & $	203	$ & $	429	$ & $	368	$ \\
$ {\rm EIC4}_p $	& $	215	$ & $	6	$ & $	779	$ & $	314	$ & $	0	$ & $	686	$ & $	424	$ & $	7	$ & $	569	$ & $	704	$ & $	11	$ & $	285	$ \\
$ {\rm EIC5}_p $	& $	93	$ & $	9	$ & $	898	$ & $	97	$ & $	11	$ & $	892	$ & $	505	$ & $	53	$ & $	442	$ & $	821	$ & $	103	$ & $	76	$ \\
$ {\rm EIC1}_{np} $	& $	991	$ & $	9	$ & $	0	$ & $	955	$ & $	36	$ & $	9	$ & $	799	$ & $	183	$ & $	18	$ & $	486	$ & $	425	$ & $	89	$ \\
$ {\rm EIC2}_{np} $	& $	997	$ & $	3	$ & $	0	$ & $	985	$ & $	11	$ & $	4	$ & $	921	$ & $	76	$ & $	3	$ & $	675	$ & $	300	$ & $	25	$ \\
$ {\rm EIC3}_{np} $	& $	463	$ & $	133	$ & $	404	$ & $	438	$ & $	273	$ & $	289	$ & $	275	$ & $	399	$ & $	326	$ & $	174	$ & $	433	$ & $	393	$ \\
$ {\rm EIC4}_{np} $	& $	998	$ & $	2	$ & $	0	$ & $	981	$ & $	15	$ & $	4	$ & $	894	$ & $	99	$ & $	7	$ & $	674	$ & $	293	$ & $	33	$ \\
$ {\rm EIC5}_{np} $	& $	281	$ & $	78	$ & $	641	$ & $	379	$ & $	243	$ & $	378	$ & $	229	$ & $	355	$ & $	416	$ & $	151	$ & $	365	$ & $	484	$ \\
BCV	& $	999	$ & $	1	$ & $	0	$ & $	993	$ & $	6	$ & $	1	$ & $	948	$ & $	50	$ & $	2	$ & $	795	$ & $	193	$ & $	12	$ \\
632CV	& $	997	$ & $	3	$ & $	0	$ & $	978	$ & $	17	$ & $	5	$ & $	890	$ & $	104	$ & $	6	$ & $	649	$ & $	308	$ & $	43	$ \\

\hline
\end{tabularx}
\end{center}}
\end{table}

\begin{table}[t]
\tablesize

\caption{Frequencies of correct and incorrect order selection from 1000 independent replications; mean and dispersion regressors jointly selected in a weakly identified model (Model (\ref{E:model8})).  
} \label{T:M1_b4}
{
\begin{center}
\begin{tabularx}{\textwidth}{ l|XXX|XXX|XXX|XXX}	
\hline
 & \multicolumn{3}{c|}{$n=25$} & \multicolumn{3}{c|}{$n=30$} & \multicolumn{3}{c|}{$n=40$} & \multicolumn{3}{c}{$n=50$}\\

	& $	<k_0	$ & $	=k_0	$ & $	>k_0	$ & $	<k_0	$ & $	=k_0	$ & $	>k_0	$ & $	<k_0	$ & $	=k_0	$ & $	>k_0	$ & $	<k_0	$ & $	=k_0	$ & $	>k_0	$ \\
\hline

AIC	& $	317	$ & $	69	$ & $	614	$ & $	439	$ & $	100	$ & $	461	$ & $	517	$ & $	136	$ & $	347	$ & $	484	$ & $	157	$ & $	359	$ \\
AICc	& $	778	$ & $	75	$ & $	147	$ & $	748	$ & $	118	$ & $	134	$ & $	736	$ & $	112	$ & $	152	$ & $	676	$ & $	150	$ & $	174	$ \\
SIC	& $	662	$ & $	56	$ & $	282	$ & $	778	$ & $	78	$ & $	144	$ & $	861	$ & $	65	$ & $	74	$ & $	889	$ & $	67	$ & $	44	$ \\
SICc	& $	963	$ & $	18	$ & $	19	$ & $	957	$ & $	33	$ & $	10	$ & $	966	$ & $	25	$ & $	9	$ & $	957	$ & $	30	$ & $	13	$ \\
HQ	& $	424	$ & $	71	$ & $	505	$ & $	572	$ & $	101	$ & $	327	$ & $	685	$ & $	105	$ & $	210	$ & $	694	$ & $	133	$ & $	173	$ \\
HQc	& $	867	$ & $	58	$ & $	75	$ & $	856	$ & $	86	$ & $	58	$ & $	867	$ & $	74	$ & $	59	$ & $	849	$ & $	95	$ & $	56	$ \\
BQCV	& $	866	$ & $	82	$ & $	52	$ & $	791	$ & $	145	$ & $	64	$ & $	699	$ & $	161	$ & $	140	$ & $	544	$ & $	\mathbf{229}	$ & $	227	$ \\
632QCV	& $	726	$ & $	\mathbf{158}	$ & $	116	$ & $	573	$ & $	\mathbf{237}	$ & $	190	$ & $	452	$ & $	\mathbf{218}	$ & $	330	$ & $	313	$ & $	225	$ & $	462	$ \\
$ {\rm EIC1}_p $	& $	960	$ & $	30	$ & $	10	$ & $	931	$ & $	50	$ & $	19	$ & $	901	$ & $	61	$ & $	38	$ & $	827	$ & $	113	$ & $	60	$ \\
$ {\rm EIC2}_p $	& $	984	$ & $	14	$ & $	2	$ & $	966	$ & $	26	$ & $	8	$ & $	949	$ & $	33	$ & $	18	$ & $	887	$ & $	80	$ & $	33	$ \\
$ {\rm EIC3}_p $	& $	885	$ & $	88	$ & $	27	$ & $	721	$ & $	213	$ & $	66	$ & $	657	$ & $	154	$ & $	189	$ & $	587	$ & $	180	$ & $	233	$ \\
$ {\rm EIC4}_p $	& $	326	$ & $	7	$ & $	667	$ & $	484	$ & $	8	$ & $	508	$ & $	641	$ & $	15	$ & $	344	$ & $	804	$ & $	44	$ & $	152	$ \\
$ {\rm EIC5}_p $	& $	11	$ & $	0	$ & $	989	$ & $	29	$ & $	0	$ & $	971	$ & $	228	$ & $	9	$ & $	763	$ & $	630	$ & $	128	$ & $	242	$ \\
$ {\rm EIC1}_{np} $	& $	994	$ & $	6	$ & $	0	$ & $	977	$ & $	17	$ & $	6	$ & $	951	$ & $	42	$ & $	7	$ & $	856	$ & $	117	$ & $	27	$ \\
$ {\rm EIC2}_{np} $	& $	1000	$ & $	0	$ & $	0	$ & $	995	$ & $	5	$ & $	0	$ & $	977	$ & $	23	$ & $	0	$ & $	930	$ & $	62	$ & $	8	$ \\
$ {\rm EIC3}_{np} $	& $	593	$ & $	91	$ & $	316	$ & $	672	$ & $	160	$ & $	168	$ & $	636	$ & $	172	$ & $	192	$ & $	582	$ & $	169	$ & $	249	$ \\
$ {\rm EIC4}_{np} $	& $	999	$ & $	1	$ & $	0	$ & $	994	$ & $	4	$ & $	2	$ & $	968	$ & $	32	$ & $	0	$ & $	912	$ & $	76	$ & $	12	$ \\
$ {\rm EIC5}_{np} $	& $	362	$ & $	34	$ & $	604	$ & $	588	$ & $	148	$ & $	264	$ & $	572	$ & $	208	$ & $	220	$ & $	496	$ & $	204	$ & $	300	$ \\
BCV	& $	1000	$ & $	0	$ & $	0	$ & $	1000	$ & $	0	$ & $	0	$ & $	991	$ & $	9	$ & $	0	$ & $	969	$ & $	28	$ & $	3	$ \\
632CV	& $	999	$ & $	1	$ & $	0	$ & $	994	$ & $	6	$ & $	0	$ & $	975	$ & $	25	$ & $	0	$ & $	911	$ & $	76	$ & $	13	$ \\

\hline
\end{tabularx}
\end{center}}
\end{table}

\begin{table}[t]
\tablesize

\caption{Frequencies of correct and incorrect order selection from 1000 independent replications;
mean regressors selected in an easily identified model (Model (\ref{E:model7})).
} \label{T:M3_b3}
{
\begin{center}
\begin{tabularx}{\textwidth}{ l|XXX|XXX|XXX|XXX}	
\hline
 & \multicolumn{3}{c|}{$n=25$} & \multicolumn{3}{c|}{$n=30$} & \multicolumn{3}{c|}{$n=40$} & \multicolumn{3}{c}{$n=50$}\\

	& $	<r_0	$ & $	=r_0	$ & $	>r_0	$ & $	<r_0	$ & $	=r_0	$ & $	>r_0	$ & $	<r_0	$ & $	=r_0	$ & $	>r_0	$ & $	<r_0	$ & $	=r_0	$ & $	>r_0	$ \\
\hline																									
AIC	& $	120	$ & $	360	$ & $	520	$ & $	98	$ & $	426	$ & $	476	$ & $	70	$ & $	531	$ & $	399	$ & $	44	$ & $	617	$ & $	339	$ \\
AICc	& $	326	$ & $	519	$ & $	155	$ & $	209	$ & $	589	$ & $	202	$ & $	114	$ & $	654	$ & $	232	$ & $	68	$ & $	739	$ & $	193	$ \\
SIC	& $	278	$ & $	461	$ & $	261	$ & $	233	$ & $	538	$ & $	229	$ & $	192	$ & $	652	$ & $	156	$ & $	144	$ & $	754	$ & $	102	$ \\
SICc	& $	559	$ & $	390	$ & $	51	$ & $	459	$ & $	486	$ & $	55	$ & $	335	$ & $	613	$ & $	52	$ & $	239	$ & $	714	$ & $	47	$ \\
HQ	& $	160	$ & $	402	$ & $	438	$ & $	136	$ & $	496	$ & $	368	$ & $	105	$ & $	607	$ & $	288	$ & $	75	$ & $	722	$ & $	203	$ \\
HQc	& $	383	$ & $	501	$ & $	116	$ & $	291	$ & $	579	$ & $	130	$ & $	190	$ & $	677	$ & $	133	$ & $	121	$ & $	769	$ & $	110	$ \\
BQCV	& $	487	$ & $	510	$ & $	3	$ & $	279	$ & $	699	$ & $	22	$ & $	115	$ & $	\mathbf{776}	$ & $	109	$ & $	56	$ & $	801	$ & $	143	$ \\
632QCV	& $	316	$ & $	\mathbf{668}	$ & $	16	$ & $	168	$ & $	\mathbf{779} $ & $	53	$ & $	65	$ & $	705	$ & $	230	$ & $	27	$ & $	673	$ & $	300	$ \\
$ {\rm EIC1}_p $	& $	932	$ & $	68	$ & $	0	$ & $	904	$ & $	94	$ & $	2	$ & $	869	$ & $	110	$ & $	21	$ & $	827	$ & $	157	$ & $	16	$ \\
$ {\rm EIC2}_p $	& $	790	$ & $	210	$ & $	0	$ & $	560	$ & $	438	$ & $	2	$ & $	297	$ & $	680	$ & $	23	$ & $	147	$ & $	\mathbf{823}	$ & $	30	$ \\
$ {\rm EIC3}_p $	& $	543	$ & $	456	$ & $	1	$ & $	279	$ & $	694	$ & $	27	$ & $	112	$ & $	705	$ & $	183	$ & $	58	$ & $	737	$ & $	205	$ \\
$ {\rm EIC4}_p $	& $	404	$ & $	4	$ & $	592	$ & $	408	$ & $	5	$ & $	587	$ & $	818	$ & $	8	$ & $	174	$ & $	959	$ & $	17	$ & $	24	$ \\
$ {\rm EIC5}_p $	& $	313	$ & $	2	$ & $	685	$ & $	722	$ & $	4	$ & $	274	$ & $	968	$ & $	9	$ & $	23	$ & $	968	$ & $	13	$ & $	19	$ \\
$ {\rm EIC1}_{np} $	& $	970	$ & $	30	$ & $	0	$ & $	927	$ & $	72	$ & $	1	$ & $	603	$ & $	384	$ & $	13	$ & $	300	$ & $	634	$ & $	66	$ \\
$ {\rm EIC2}_{np} $	& $	974	$ & $	26	$ & $	0	$ & $	958	$ & $	41	$ & $	1	$ & $	725	$ & $	268	$ & $	7	$ & $	471	$ & $	495	$ & $	34	$ \\
$ {\rm EIC3}_{np} $	& $	325	$ & $	502	$ & $	173	$ & $	222	$ & $	624	$ & $	154	$ & $	126	$ & $	677	$ & $	197	$ & $	76	$ & $	722	$ & $	202	$ \\
$ {\rm EIC4}_{np} $	& $	974	$ & $	26	$ & $	0	$ & $	956	$ & $	43	$ & $	1	$ & $	723	$ & $	269	$ & $	8	$ & $	463	$ & $	497	$ & $	40	$ \\
$ {\rm EIC5}_{np} $	& $	324	$ & $	426	$ & $	250	$ & $	248	$ & $	558	$ & $	194	$ & $	162	$ & $	584	$ & $	254	$ & $	83	$ & $	616	$ & $	301	$ \\
BCV	& $	976	$ & $	24	$ & $	0	$ & $	965	$ & $	35	$ & $	0	$ & $	783	$ & $	210	$ & $	7	$ & $	582	$ & $	396	$ & $	22	$ \\
632CV	& $	975	$ & $	25	$ & $	0	$ & $	953	$ & $	47	$ & $	0	$ & $	689	$ & $	302	$ & $	9	$ & $	419	$ & $	540	$ & $	41	$ \\

\hline
\end{tabularx}
\end{center}}
\end{table}

\begin{table}[t]
\tablesize

\caption{Frequencies of correct and incorrect order selection from 1000 independent replications; 
mean regressors selected in a weakly identified model (Model (\ref{E:model8})). 
} \label{T:M3_b4}
{
\begin{center}
\begin{tabularx}{\textwidth}{ l|XXX|XXX|XXX|XXX}	
\hline
 & \multicolumn{3}{c|}{$n=25$} & \multicolumn{3}{c|}{$n=30$} & \multicolumn{3}{c|}{$n=40$} & \multicolumn{3}{c}{$n=50$}\\

	& $	<r_0	$ & $	=r_0	$ & $	>r_0	$ & $	<r_0	$ & $	=r_0	$ & $	>r_0	$ & $	<r_0	$ & $	=r_0	$ & $	>r_0	$ & $	<r_0	$ & $	=r_0	$ & $	>r_0	$ \\
\hline																									
AIC	& $	369	$ & $	173	$ & $	458	$ & $	421	$ & $	202	$ & $	377	$ & $	437	$ & $	256	$ & $	307	$ & $	453	$ & $	269	$ & $	278	$ \\
AICc	& $	711	$ & $	171	$ & $	118	$ & $	702	$ & $	199	$ & $	99	$ & $	614	$ & $	250	$ & $	136	$ & $	598	$ & $	259	$ & $	143	$ \\
SIC	& $	626	$ & $	153	$ & $	221	$ & $	729	$ & $	164	$ & $	107	$ & $	722	$ & $	196	$ & $	82	$ & $	772	$ & $	180	$ & $	48	$ \\
SICc	& $	892	$ & $	86	$ & $	22	$ & $	893	$ & $	93	$ & $	14	$ & $	850	$ & $	131	$ & $	19	$ & $	851	$ & $	135	$ & $	14	$ \\
HQ	& $	449	$ & $	173	$ & $	378	$ & $	535	$ & $	204	$ & $	261	$ & $	572	$ & $	246	$ & $	182	$ & $	611	$ & $	246	$ & $	143	$ \\
HQc	& $	783	$ & $	145	$ & $	72	$ & $	791	$ & $	152	$ & $	57	$ & $	724	$ & $	204	$ & $	72	$ & $	727	$ & $	211	$ & $	62	$ \\
BQCV	& $	846	$ & $	153	$ & $	1	$ & $	779	$ & $	211	$ & $	10	$ & $	600	$ & $	330	$ & $	70	$ & $	546	$ & $	\mathbf{325}	$ & $	129	$ \\
632QCV	& $	743	$ & $	\mathbf{247}	$ & $	10	$ & $	651	$ & $	\mathbf{307}	$ & $	42	$ & $	451	$ & $	\mathbf{376}	$ & $	173	$ & $	378	$ & $	323	$ & $	299	$ \\
$ {\rm EIC1}_p $	& $	941	$ & $	58	$ & $	1	$ & $	908	$ & $	90	$ & $	2	$ & $	836	$ & $	139	$ & $	25	$ & $	796	$ & $	168	$ & $	36	$ \\
$ {\rm EIC2}_p $	& $	958	$ & $	42	$ & $	0	$ & $	918	$ & $	82	$ & $	0	$ & $	808	$ & $	182	$ & $	10	$ & $	765	$ & $	221	$ & $	14	$ \\
$ {\rm EIC3}_p $	& $	856	$ & $	142	$ & $	2	$ & $	737	$ & $	243	$ & $	20	$ & $	582	$ & $	297	$ & $	121	$ & $	551	$ & $	288	$ & $	161	$ \\
$ {\rm EIC4}_p $	& $	583	$ & $	17	$ & $	400	$ & $	625	$ & $	21	$ & $	354	$ & $	848	$ & $	60	$ & $	92	$ & $	899	$ & $	75	$ & $	26	$ \\
$ {\rm EIC5}_p $	& $	52	$ & $	1	$ & $	947	$ & $	238	$ & $	4	$ & $	758	$ & $	764	$ & $	89	$ & $	147	$ & $	796	$ & $	101	$ & $	103	$ \\
$ {\rm EIC1}_{np} $	& $	982	$ & $	17	$ & $	1	$ & $	985	$ & $	15	$ & $	0	$ & $	905	$ & $	93	$ & $	2	$ & $	827	$ & $	166	$ & $	7	$ \\
$ {\rm EIC2}_{np} $	& $	983	$ & $	16	$ & $	1	$ & $	988	$ & $	12	$ & $	0	$ & $	945	$ & $	54	$ & $	1	$ & $	875	$ & $	123	$ & $	2	$ \\
$ {\rm EIC3}_{np} $	& $	696	$ & $	175	$ & $	129	$ & $	717	$ & $	218	$ & $	65	$ & $	618	$ & $	278	$ & $	104	$ & $	576	$ & $	290	$ & $	134	$ \\
$ {\rm EIC4}_{np} $	& $	983	$ & $	16	$ & $	1	$ & $	990	$ & $	10	$ & $	0	$ & $	940	$ & $	59	$ & $	1	$ & $	873	$ & $	124	$ & $	3	$ \\
$ {\rm EIC5}_{np} $	& $	652	$ & $	149	$ & $	199	$ & $	690	$ & $	219	$ & $	91	$ & $	579	$ & $	300	$ & $	121	$ & $	549	$ & $	292	$ & $	159	$ \\
BCV	& $	985	$ & $	14	$ & $	1	$ & $	992	$ & $	8	$ & $	0	$ & $	958	$ & $	41	$ & $	1	$ & $	907	$ & $	91	$ & $	2	$ \\
632CV	& $	985	$ & $	14	$ & $	1	$ & $	989	$ & $	11	$ & $	0	$ & $	934	$ & $	65	$ & $	1	$ & $	863	$ & $	133	$ & $	4	$ \\

\hline
\end{tabularx}
\end{center}}
\end{table}

\begin{table}[t]
\tablesize

\caption{Frequencies of correct and incorrect order selection from 1000 independent replications; dispersion regressors selected in an easily identified model (Model (\ref{E:model7})).
} \label{T:M4_b3}
{
\begin{center}
\begin{tabularx}{\textwidth}{ l|XXX|XXX|XXX|XXX}	
\hline
 & \multicolumn{3}{c|}{$n=25$} & \multicolumn{3}{c|}{$n=30$} & \multicolumn{3}{c|}{$n=40$} & \multicolumn{3}{c}{$n=50$}\\

	& $	<s_0	$ & $	=s_0	$ & $	>s_0	$ & $	<s_0	$ & $	=s_0	$ & $	>s_0	$ & $	<s_0	$ & $	=s_0	$ & $	>s_0	$ & $	<s_0	$ & $	=s_0	$ & $	>s_0	$ \\
\hline																									
AIC	& $	328	$ & $	230	$ & $	442	$ & $	295	$ & $	313	$ & $	392	$ & $	254	$ & $	401	$ & $	345	$ & $	246	$ & $	466	$ & $	288	$ \\
AICc	& $	632	$ & $	252	$ & $	116	$ & $	523	$ & $	342	$ & $	135	$ & $	396	$ & $	452	$ & $	152	$ & $	339	$ & $	499	$ & $	162	$ \\
SIC	& $	567	$ & $	221	$ & $	212	$ & $	553	$ & $	303	$ & $	144	$ & $	519	$ & $	394	$ & $	87	$ & $	535	$ & $	409	$ & $	56	$ \\
SICc	& $	838	$ & $	143	$ & $	19	$ & $	785	$ & $	183	$ & $	32	$ & $	679	$ & $	297	$ & $	24	$ & $	650	$ & $	331	$ & $	19	$ \\
HQ	& $	398	$ & $	247	$ & $	355	$ & $	378	$ & $	332	$ & $	290	$ & $	361	$ & $	430	$ & $	209	$ & $	357	$ & $	480	$ & $	163	$ \\
HQc	& $	705	$ & $	223	$ & $	72	$ & $	620	$ & $	301	$ & $	79	$ & $	508	$ & $	421	$ & $	71	$ & $	467	$ & $	465	$ & $	68	$ \\
BQCV	& $	882	$ & $	118	$ & $	0	$ & $	783	$ & $	216	$ & $	1	$ & $	471	$ & $	493	$ & $	36	$ & $	348	$ & $	571	$ & $	81	$ \\
632QCV	& $	734	$ & $	\mathbf{266}	$ & $	0	$ & $	574	$ & $	\mathbf{413}	$ & $	13	$ & $	304	$ & $	\mathbf{582}	$ & $	114	$ & $	214	$ & $	\mathbf{572}	$ & $	214	$ \\
$ {\rm EIC1}_p $	& $	974	$ & $	26	$ & $	0	$ & $	941	$ & $	59	$ & $	0	$ & $	810	$ & $	174	$ & $	16	$ & $	716	$ & $	227	$ & $	57	$ \\
$ {\rm EIC2}_p $	& $	994	$ & $	6	$ & $	0	$ & $	969	$ & $	31	$ & $	0	$ & $	800	$ & $	196	$ & $	4	$ & $	664	$ & $	327	$ & $	9	$ \\
$ {\rm EIC3}_p $	& $	842	$ & $	158	$ & $	0	$ & $	625	$ & $	371	$ & $	4	$ & $	348	$ & $	511	$ & $	141	$ & $	299	$ & $	517	$ & $	184	$ \\
$ {\rm EIC4}_p $	& $	617	$ & $	21	$ & $	362	$ & $	622	$ & $	15	$ & $	363	$ & $	701	$ & $	55	$ & $	244	$ & $	809	$ & $	116	$ & $	75	$ \\
$ {\rm EIC5}_p $	& $	155	$ & $	11	$ & $	834	$ & $	313	$ & $	37	$ & $	650	$ & $	541	$ & $	140	$ & $	319	$ & $	535	$ & $	174	$ & $	291	$ \\
$ {\rm EIC1}_{np} $	& $	1000	$ & $	0	$ & $	0	$ & $	995	$ & $	5	$ & $	0	$ & $	870	$ & $	127	$ & $	3	$ & $	736	$ & $	246	$ & $	18	$ \\
$ {\rm EIC2}_{np} $	& $	1000	$ & $	0	$ & $	0	$ & $	999	$ & $	1	$ & $	0	$ & $	952	$ & $	47	$ & $	1	$ & $	883	$ & $	114	$ & $	3	$ \\
$ {\rm EIC3}_{np} $	& $	564	$ & $	236	$ & $	200	$ & $	509	$ & $	355	$ & $	136	$ & $	358	$ & $	468	$ & $	174	$ & $	290	$ & $	480	$ & $	230	$ \\
$ {\rm EIC4}_{np} $	& $	1000	$ & $	0	$ & $	0	$ & $	1000	$ & $	0	$ & $	0	$ & $	947	$ & $	52	$ & $	1	$ & $	872	$ & $	126	$ & $	2	$ \\
$ {\rm EIC5}_{np} $	& $	533	$ & $	183	$ & $	284	$ & $	514	$ & $	303	$ & $	183	$ & $	382	$ & $	388	$ & $	230	$ & $	318	$ & $	388	$ & $	294	$ \\
BCV	& $	1000	$ & $	0	$ & $	0	$ & $	1000	$ & $	0	$ & $	0	$ & $	962	$ & $	38	$ & $	0	$ & $	922	$ & $	76	$ & $	2	$ \\
632CV	& $	1000	$ & $	0	$ & $	0	$ & $	998	$ & $	2	$ & $	0	$ & $	935	$ & $	62	$ & $	3	$ & $	852	$ & $	145	$ & $	3	$ \\

\hline
\end{tabularx}
\end{center}}
\end{table}

\begin{table}[t]
\tablesize

\caption{Frequencies of correct and incorrect order selection from 1000 independent replications; dispersion regressors selected in a weakly identified model (Model (\ref{E:model8})).
} \label{T:M4_b4}
{
\begin{center}
\begin{tabularx}{\textwidth}{ l|XXX|XXX|XXX|XXX}	
\hline
 & \multicolumn{3}{c|}{$n=25$} & \multicolumn{3}{c|}{$n=30$} & \multicolumn{3}{c|}{$n=40$} & \multicolumn{3}{c}{$n=50$}\\

	& $	<s_0	$ & $	=s_0	$ & $	>s_0	$ & $	<s_0	$ & $	=s_0	$ & $	>s_0	$ & $	<s_0	$ & $	=s_0	$ & $	>s_0	$ & $	<s_0	$ & $	=s_0	$ & $	>s_0	$ \\
\hline																									
AIC	& $	455	$ & $	110	$ & $	435	$ & $	487	$ & $	156	$ & $	357	$ & $	520	$ & $	201	$ & $	279	$ & $	531	$ & $	214	$ & $	255	$ \\
AICc	& $	812	$ & $	98	$ & $	90	$ & $	762	$ & $	135	$ & $	103	$ & $	689	$ & $	182	$ & $	129	$ & $	669	$ & $	206	$ & $	125	$ \\
SIC	& $	738	$ & $	89	$ & $	173	$ & $	782	$ & $	109	$ & $	109	$ & $	807	$ & $	122	$ & $	71	$ & $	818	$ & $	147	$ & $	35	$ \\
SICc	& $	947	$ & $	38	$ & $	15	$ & $	939	$ & $	50	$ & $	11	$ & $	913	$ & $	69	$ & $	18	$ & $	900	$ & $	87	$ & $	13	$ \\
HQ	& $	537	$ & $	\mathbf{114}	$ & $	349	$ & $	606	$ & $	152	$ & $	242	$ & $	658	$ & $	186	$ & $	156	$ & $	685	$ & $	192	$ & $	123	$ \\
HQc	& $	876	$ & $	70	$ & $	54	$ & $	848	$ & $	104	$ & $	48	$ & $	809	$ & $	131	$ & $	60	$ & $	787	$ & $	164	$ & $	49	$ \\
BQCV	& $	975	$ & $	25	$ & $	0	$ & $	928	$ & $	70	$ & $	2	$ & $	800	$ & $	175	$ & $	25	$ & $	690	$ & $	243	$ & $	67	$ \\
632QCV	& $	911	$ & $	88	$ & $	1	$ & $	832	$ & $	161	$ & $	7	$ & $	619	$ & $	\mathbf{290}	$ & $	91	$ & $	530	$ & $	\mathbf{293}	$ & $	177	$ \\
$ {\rm EIC1}_p $	& $	991	$ & $	9	$ & $	0	$ & $	967	$ & $	33	$ & $	0	$ & $	900	$ & $	93	$ & $	7	$ & $	833	$ & $	139	$ & $	28	$ \\
$ {\rm EIC2}_p $	& $	999	$ & $	1	$ & $	0	$ & $	997	$ & $	3	$ & $	0	$ & $	950	$ & $	49	$ & $	1	$ & $	911	$ & $	83	$ & $	6	$ \\
$ {\rm EIC3}_p $	& $	943	$ & $	57	$ & $	0	$ & $	812	$ & $	\mathbf{180}	$ & $	8	$ & $	665	$ & $	237	$ & $	98	$ & $	610	$ & $	239	$ & $	151	$ \\
$ {\rm EIC4}_p $	& $	729	$ & $	18	$ & $	253	$ & $	778	$ & $	6	$ & $	216	$ & $	829	$ & $	26	$ & $	145	$ & $	900	$ & $	78	$ & $	22	$ \\
$ {\rm EIC5}_p $	& $	45	$ & $	2	$ & $	953	$ & $	181	$ & $	16	$ & $	803	$ & $	559	$ & $	147	$ & $	294	$ & $	566	$ & $	188	$ & $	246	$ \\
$ {\rm EIC1}_{np} $	& $	1000	$ & $	0	$ & $	0	$ & $	1000	$ & $	0	$ & $	0	$ & $	967	$ & $	33	$ & $	0	$ & $	892	$ & $	101	$ & $	7	$ \\
$ {\rm EIC2}_{np} $	& $	1000	$ & $	0	$ & $	0	$ & $	1000	$ & $	0	$ & $	0	$ & $	985	$ & $	15	$ & $	0	$ & $	950	$ & $	47	$ & $	3	$ \\
$ {\rm EIC3}_{np} $	& $	723	$ & $	110	$ & $	167	$ & $	742	$ & $	151	$ & $	107	$ & $	636	$ & $	224	$ & $	140	$ & $	606	$ & $	228	$ & $	166	$ \\
$ {\rm EIC4}_{np} $	& $	1000	$ & $	0	$ & $	0	$ & $	1000	$ & $	0	$ & $	0	$ & $	987	$ & $	13	$ & $	0	$ & $	945	$ & $	51	$ & $	4	$ \\
$ {\rm EIC5}_{np} $	& $	643	$ & $	88	$ & $	269	$ & $	711	$ & $	132	$ & $	157	$ & $	595	$ & $	231	$ & $	174	$ & $	546	$ & $	262	$ & $	192	$ \\
BCV	& $	1000	$ & $	0	$ & $	0	$ & $	1000	$ & $	0	$ & $	0	$ & $	994	$ & $	6	$ & $	0	$ & $	971	$ & $	27	$ & $	2	$ \\
632CV	& $	1000	$ & $	0	$ & $	0	$ & $	1000	$ & $	0	$ & $	0	$ & $	980	$ & $	20	$ & $	0	$ & $	933	$ & $	62	$ & $	5	$ \\

\hline
\end{tabularx}
\end{center}}
\end{table}

In all cases, the correct model order dimension is $k_0=6$: there are three parameters in the mean submodel and three parameters in the regression structure for the dispersion. The sample sizes are $n=25, 30, 40, 50$ and five candidate covariates are considered for both submodels. The candidate models are sequentially nested for the mean submodel, that is, the candidate model with $r$ parameters in the mean regression structure consists of the submodel with the $1,2,\dots,r$ first parameters. The dispersion submodels are also sequentially nested. Thus, for each value of $r$ we vary $s$ from $1$ to $6$, totaling $6 \times 6 = 36$ candidate models.

Since the true model belongs to the set of candidate models, the  evaluation of the different selection criteria is done by counting the number of times that each criterion selects the correct model order ($k_0$, $r_0$ or $s_0$). Three different approaches were considered. First, we used the different model selection criteria to jointly select the mean and dispersion regressors; the results are given in Tables~\ref{T:M1_b3} and \ref{T:M1_b4}. 
Afterwards, for a correctly specified dispersion submodel, we used the model selection criteria to select the regressors in the mean submodel; the results are given in Tables~\ref{T:M3_b3} and \ref{T:M3_b4}. Finally, for a correctly specified mean submodel, we performed model selection on the dispersion submodel; the results are presented in Tables~\ref{T:M4_b3} and \ref{T:M4_b4}. In all tables the best results are highlighted. 

The figures in Table~\ref{T:M1_b3} show the proposed criteria yield reliable joint selection of mean and dispersion regressors in easily identifiable models. We note that for $n=25$ and $n=30$ 632QCV was the best performing criterion. For $n=40$ and $n=50$ BQCV was the best performer.  
Among the extensions (EIC's) of the AIC, the criterion that stands out is the EIC3 in their two versions, both with parametric and with nonparametric bootstrap. In this scenario, the AICc stands out when compared to alternative criteria that do not make use of bootstrapped log-likelihood. It is noteworthy the poor performance of the BCV, 632CV and EIC's criteria. When the sample size increases, the performances of the nonparametric EIC's improve, becoming similar. The same does not hold, however, for the parametric EIC's: $ {\rm EIC1}_p $ and ${\rm EIC4}_p$ perform poorly in all sample sizes. 

Under weak identifiability the good performances of the BQCV and 632QCV criteria become even more evident; see Table~\ref{T:M1_b4}. The 632QCV criterion was the best performer for $n=25, 30, 40$. For $n=50$, BQCV outperformed the competition. It is noteworthy that for $n=25, 30$ the 632QCV criterion outperformed all nonbootstrap based criteria by at least $200\%$. The EIC3 performs well relative to the other bootstrap extensions when regressors are jointly selected for both submodels in a weakly identifiable model. We also note the weak performances of the BCV and 632CV criteria. The AICc clearly outperforms the AIC. For instance, the AIC selected an overspecified model in 614 replications whereas that happened only 147 times when the AICc was used.


We shall now focus on selecting regressors for the mean submodel. Here, the dispersion submodel is correctly specified and the interest lies in identifying which covariates must be included in the mean submodel. The results for a weakly identifiable model are displayed in Table~\ref{T:M3_b3}. They again show the good finite sample performances of our two model selection criteria. For $n=25,30$, the 632QCV criterion was the best performer. For $n=40$, the best performer was BQCV, and for $n=50$ the ${\rm EIC2}_p$ criterion outperformed the competition. Once again, the best performing AIC extension was EIC3 and the BCV and 632CV criteria performed poorly. The figures in Table \ref{T:M3_b3} also show that the AICc and the HQc are the best performers among the criteria that do not use bootstrapped log-likelihood.

Table~\ref{T:M3_b4} contains the frequencies of correct model selection for the mean submodel when the model is weakly identifiable. The criteria that stands out are the same of the previous settings. For $n=25,30,40$ ($n=50$) 632QCV (BQCV) was the best performer. The $ {\rm EIC5}_p $ criterion tends to select models that are overspecified in small samples; see also Table~\ref{T:M1_b4}. 

In our third and final approach, the mean submodel is correctly specified and the interest lies in selecting covariates for the dispersion submodel. 
The results are presented in Table~\ref{T:M4_b3}. They show that the 632QCV criterion performs well when the model is easily identifiable; indeed, it was the best performer in all sample sizes. The 632QCV criterion was the only bootstrap AIC variant that outperformed all nonbootstrap-based criteria when $n=25$. For the remaining sample sizes, only BQCV and ${\rm EIC3}_p$ outperformed the criteria that do not employ bootstrapped log-likelihood.
Table~\ref{T:M4_b4} presents results for a  
weakly identifiable model. This was the only scenario in which 632QCV was not the best performing model selection criterion for $n=25,30$; it still performs well, nonetheless. For larger sample sizes, $n=40,50$, the proposed criterion was the best performer. For $n=25$ ($n=50$), model selection based on the HQ (${\rm EIC3}_p$) criterion was the most accurate. 

The simulation results presented above lead to important conclusions on 
beta regression model selection. Such conclusions can be summarized as follows:  
\begin{itemize}
\item The model selection criteria proposed in this paper generally work very well and lead to accurate model selection. The 632QCV criterion performed better was the sample size was small and the BQCV performed better in larger samples. 
\item Among the criteria that do not use the bootstrapped log-likelihood, the AICc and the HQc criteria were the best performers. The AICc stood out when the sample size was small and the HQc performed better in larger samples. 
\item Among the AIC extensions (EIC's), the EIC3 was the criterion that delivered most accurate model selection. Its nonparametric bootstrap implementation ($ {\rm EIC3}_{np} $) displayed the best performances in small samples and ${\rm EIC3}_p$ performed best in larger sample sizes.
\item The finite sample performances of the different information criteria are considerably superior when such criteria are used to select regressors for the mean submodel relative to dispersion submodel selection; compare the results in Tables~\ref{T:M3_b3} and Table \ref{T:M4_b3}, and also the results in Tables~\ref{T:M3_b4} and \ref{T:M4_b4}. 
\item The criteria that employ bootstrapped log-likelihood for beta regression model selection clearly outperform the competition. 
\end{itemize}

\section{Application}\label{S:application_BQCV}

We use the data given in \citeauthor{Griffiths1993} (\citeyear{Griffiths1993}, Table~15.4) on food expenditure, income and number of people in 38 households of a major city in the United States. These data were modeled by \cite{Ferrari2004}, who used a constant dispersion beta regression. We performed model selection using the two-step model selection scheme proposed in \cite{Bayer2014a} coupled with the BQCV and 632QCV criteria proposed in this paper. In this scheme, the dispersion is taken to be constant and the mean submodel covariates are selected; next, using the selected mean submodel, model selection is carried out in the dispersion submodel. As shown in \cite{Bayer2014a}, this selection scheme tends typically outperforms the joint selection of regressors for the the mean and dispersion submodels at a much lower computational cost.
An implementation of such a model selection procedure in {\tt R} language~\citep{R2014} with the proposed BQCV and 632QCV criteria and
two-step scheme is available at \url{http://www.ufsm.br/bayer/auto-beta-reg.zip}. 
The file contains computer code for model selection in beta regressions and also the dataset used in this empirical application. 

Following \cite{Ferrari2004}, we model the proportion of food expenditure $(y)$ as a function of income $(x_2)$ and of the number of people $(x_3)$ in each household. We use the logit link function for the mean and dispersion submodels. We also include as candidate covariates for both submodels the interaction between the income and the number of people $(x_4=x_2 \times x_3)$, income squared $(x_5=x_2^2)$ and a quadratic transformation of $x_3$, that is, $x_6=x_3^2$. 

Assuming constant dispersion, the selected mean submodel, both by BQCV and by 632QCV, use $x_3$ and $x_4$ as covariates. Assuming that this is the correct submodel for mean we now select the regressors to be included in the dispersion submodel. The dispersion submodel selected by the BQCV and 632QCV criteria only includes one covariate, namely: $x_3$. The parameter estimates of the selected model are presented in the Table \ref{T:food}.

\begin{table}[!h]	
\tablesize
\caption{Parameter estimates of the selected varying dispersion beta regression model; data on food expenditure.
} \label{T:food}																								
\begin{center}																									
\begin{tabular}{lrrrr}																									
\hline
Parameter	& 	Estimate	 & 	Std. error &	$z$ stat & $p$-value \\
\hline																									
\multicolumn{5}{c}{submodel for $\mu$}\\
\hline
$\beta_1$ (Constant) & $-1.3040$ & $0.1103$ & $-11.826$ & $0.0000$ \\
$\beta_3$ (Number of people) & $0.2890$ & $0.0754$ & $3.835$ & $0.0005$ \\
$\beta_4$ (Interaction) & $-0.0031$ & $0.0011$ & $-2.975$ & $0.0054$ \\
\hline																									
\multicolumn{5}{c}{submodel for $\sigma$}\\
\hline 
$\gamma_1$ (Constant) & $-2.4825$ & $0.3720$ & $-6.673$ & $0.0000$ \\
$\gamma_3$ (Number of people) & $0.2011$ & $0.1118$ & $1.798$ & $0.0813$ \\

\hline 
\multicolumn{5}{c}{$R^2_{FC}=0.4586 $} \\ 
\multicolumn{5}{c}{$R^2_{LR}=0.5448 $}\\

\hline																		
\end{tabular}																									
\end{center}																									
\end{table}

\begin{figure}[!ht]
\centering
\subfigure[Residuals vs.\ index]{\label{F:food_diag_a} \includegraphics[width=0.48\textwidth]{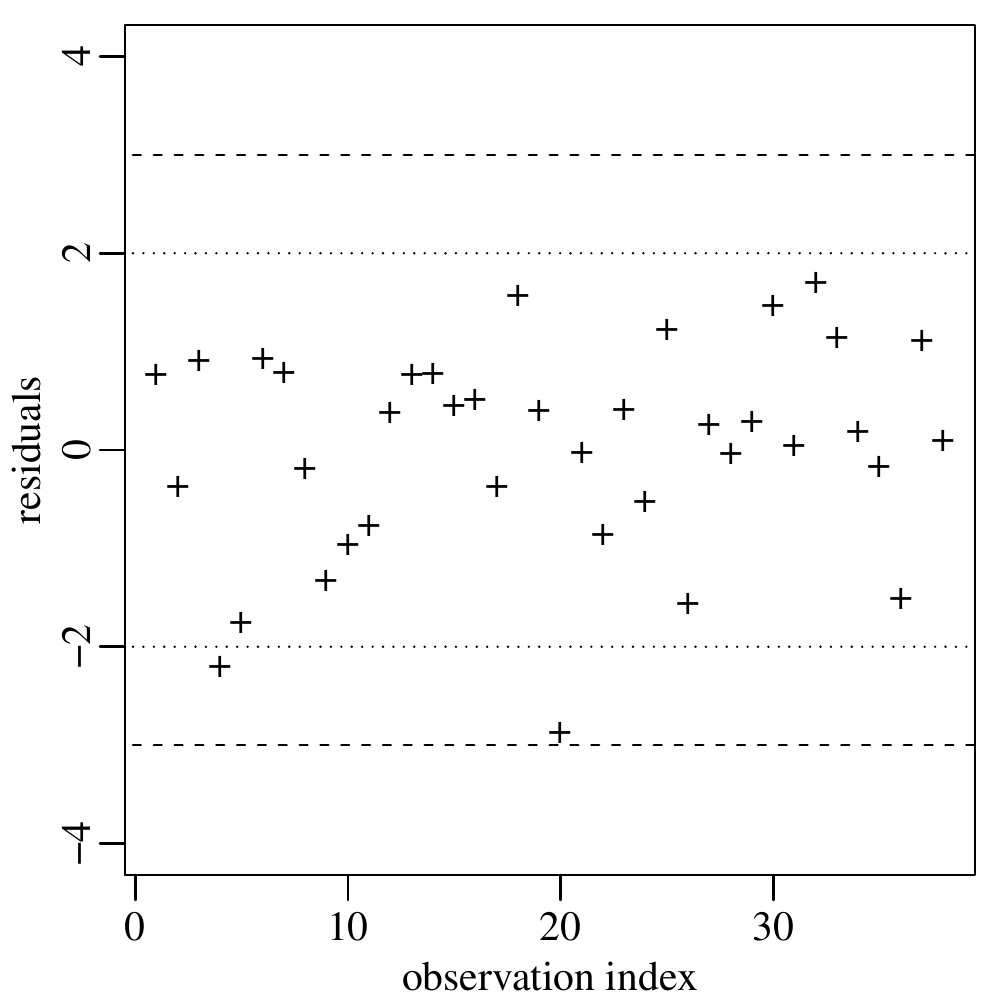}}
\subfigure[Residuals vs.\ fitted values]{\label{F:food_diag_b} \includegraphics[width=0.48\textwidth]{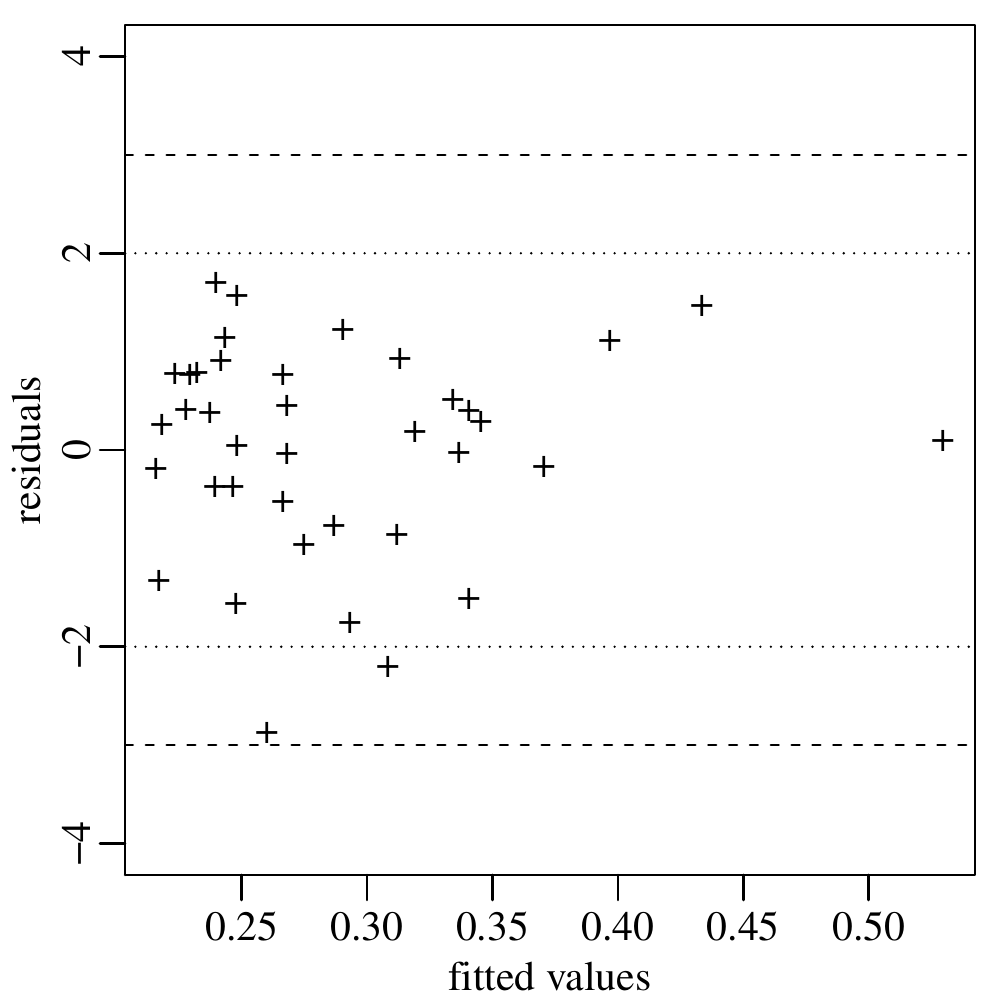}}
\subfigure[Observed values vs.\ fitted values]{\label{F:food_diag_c} \includegraphics[width=0.48\textwidth]{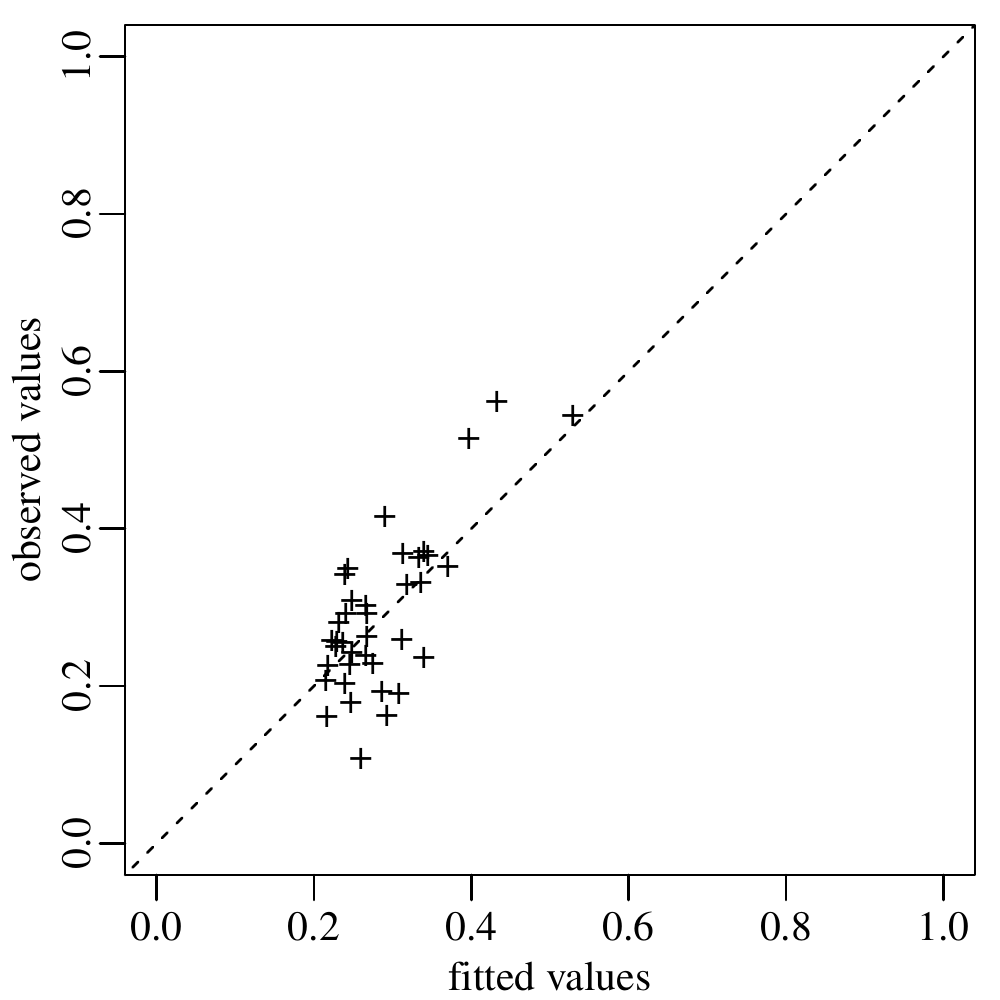}}
\subfigure[Normal probability plot]{\label{F:food_diag_d} \includegraphics[width=0.48\textwidth]{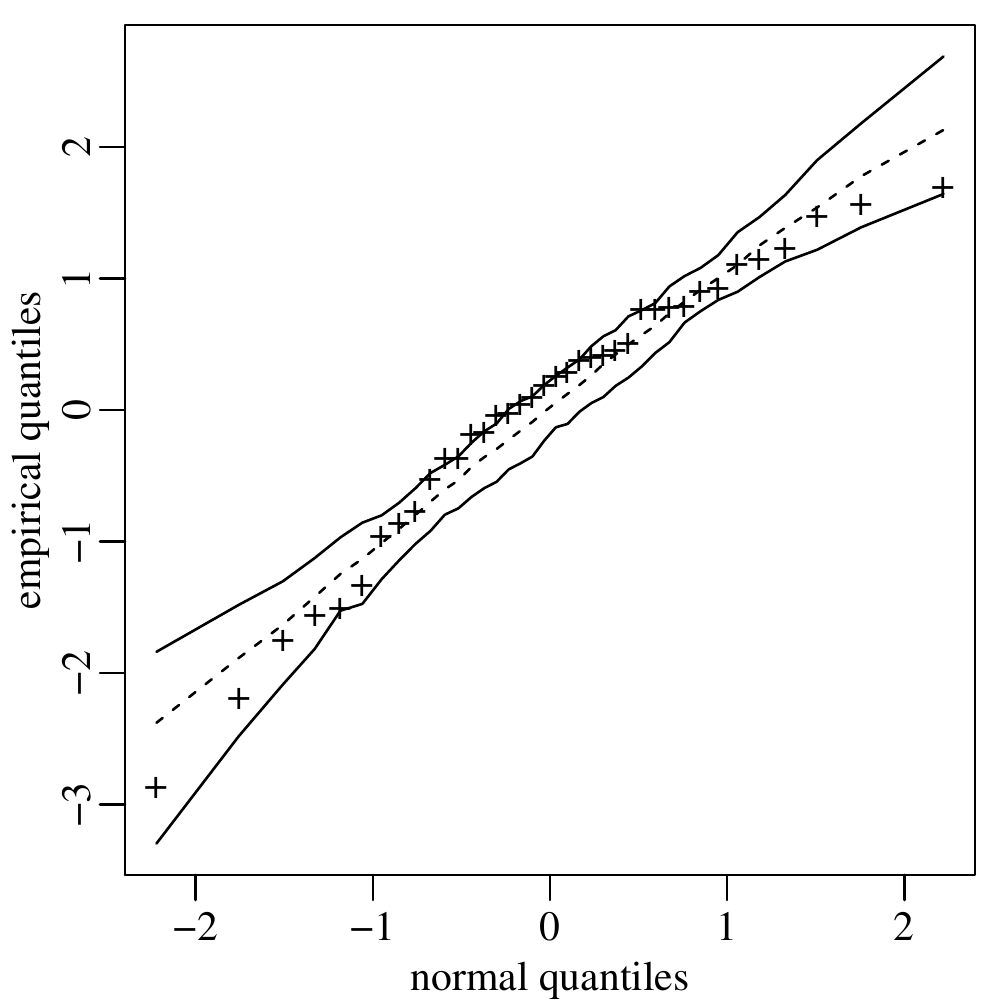}}
\caption{Residuals plots; data on food expenditure.
}\label{F:food_diag}
\end{figure}

In order to determine whether the model is correctly specified we produced residuals plots; see Figure \ref{F:food_diag}. For details on beta regression residuals and diagnostic tools, the reader is referred to \cite{Espinheira2008a,Espinheira2008b} and \cite{Ferrari2011}. We use `the standardized weighted residual 2', which is defined as 
\begin{equation*}
r_{w2} = \frac{y^{\ast}_t - \hat{\mu}^{\ast}_t}{\sqrt{\hat{v}_t(1-h_{tt})}},
\end{equation*}
where
$v^{\ast}_t= \psi' \left(\mu_t\left(\frac{1-\sigma^2_t}{\sigma^2_t} \right) \right)+ \psi' \left((1-\mu_t) \left(\frac{1-\sigma^2_t}{\sigma^2_t} \right) \right)$ and $h_{tt}$ is the $t$th diagonal element of the hat matrix (for details, see \cite{Espinheira2008b} and \cite{Ferrari2011}).

Figures~\ref{F:food_diag_a} and \ref{F:food_diag_b} show that the residuals are randomly distributed around zero. There are only two atypical observations, with residuals slightly below $-2$, but still within the range $(-3,3)$. Figure~\ref{F:food_diag_c} shows that the model fits the data well since the fitted values are similar to the observed response values. The simulated envelope plot, Figure~\ref{F:food_diag_d}, indicates that the model is correctly specified, since most points are within the envelope limits, with a few points lying on one of the envelope bands. 

We note that the parameter estimates show that there is a positive relation between the mean response and the number of people in each household, as well as a negative relationship with the interaction variable ($x_4$). There is also a positive relationship between the number of people in each household and the response dispersion. The varying dispersion beta regression model we selected and fitted has a pseudo-$R^2$ considerably larger than that of the constant dispersion model used by \cite{Ferrari2004}:  $R^2_{ML}=0.5448$ versus $R^2_{ML}=0.4088$.

\section{Conclusions}\label{S:conclusions} 


In this paper we proposed two new model selection criteria for the class of varying dispersion beta regression models. The new criteria were obtained as bootstrap variations of the AIC and provide direct estimators for the expected log-likelihood. The proposed criteria are based on the bootstrap method and on a procedure called quasi-CV. They are then called bootstrapped likelihood quasi-CV (BQCV) and 632QCV. The Monte Carlo evidence we presented favors the criteria we proposed: they typically lead to more accurate model selection than alternative criteria. We presented numerical evidence on the joint selection of regressors for the two submodels, for the mean submodel and for the dispersion submodel. An empirical application was also presented and discussed.

\section*{Acknowledgements}

We gratefully acknowledge partial financial support from CAPES, CNPq, and FAPERGS.

\appendix

\section{Score function and information matrix of the beta regression model with varying dispersion}\label{A:escore_fisher}

\small
This appendix presents the score function and the Fisher's information matrix for the beta regression model with varying dispersion presented in Section \ref{S:modelo_reg_beta}.

The score function is obtained by differentiating the log-likelihood function with respect to the unknown parameters. The score function for with respect to ${\beta}$ is given by
\begin{equation*}
U_{\!\beta}({\beta},{\gamma})= X^{\top} \Phi \, T({y}^{\ast}-{\mu}^{\ast}),
\end{equation*}
where 
$\Phi \!= \text{diag}\!
\left\{\frac{1- \sigma^2_1}{\sigma^2_1},\ldots,\frac{1-\sigma^2_n}{\sigma^2_n} \right\}$,
$T = \text{diag}\left\{\frac{1}{g^{\prime}(\mu_1)},\ldots,\frac{1}{g^{\prime}(\mu_n)}\right\}$,   ${y}^{\ast}\!=(y^{\ast}_1,\ldots,y^{\ast}_n)^{\top}$,   ${\mu}^{\ast}\!=(\mu^{\ast}_1,\ldots,\mu^{\ast}_n)^{\top}$,
$y^{\ast}_t \! = \log \left( \frac{y_t}{1-y_t} \right)$,
$\mu^{\ast}_t= \psi \left(\mu_t\left(\frac{1-\sigma^2_t}{\sigma^2_t}\right)\right)-
\psi \left((1-\mu_t)\left(\frac{1-\sigma^2_t}{\sigma^2_t}\right) \right)$
and $\psi(\cdot)$ is the digamma function, i.e., $\psi(u)\!=\frac{\partial\log\Gamma(u)}{\partial u}$, for $u>0$.
The score function with respect to ${\gamma}$ is
\begin{equation*}
U_{\!\gamma}({\beta},{\gamma})=Z^{\top}H {a},
\end{equation*}
where 
$H\! = \!\text{diag} \!\left\{\! \frac{1}{h^{\prime}(\sigma_1)}, \ldots, \frac{1}{h^{\prime}(\sigma_n)} \!\right\} $ and
${a}= ( a_1,$ $ \ldots,a_n)^{\top}$, the $t$th element of $a$ being 
$a_t = -\frac{2}{\sigma^3_t} \left\{\mu_t \left( y^*_t - \mu^*_t \right) + \log(1-y_t) - \psi \left( (1-\mu_t)(1-\sigma^2_t)/\sigma^2_t \right) + \psi\left((1-\sigma^2_t) / \sigma^2_t \right)    \right\}$.

Fisher's information matrix for ${\beta}$ and ${\gamma}$ is given by
\begin{equation}\label{E:Fisher}
K({\beta},{\gamma}) = \left(
\begin{array}{cc}
	K_{({\beta},{\beta})} & K_{({\beta},{\gamma})} \\
	K_{({\gamma},{\beta})} & K_{({\gamma},{\gamma})}
\end{array}
 \right),
\end{equation}
where $K_{({\beta},{\beta})} = X^{\top}\Phi WX $, $K_{({\beta},{\gamma})} = (K_{({\gamma},{\beta})})^{\top}=X^{\top}CTHZ$ and $K_{({\gamma},{\gamma})}=Z^{\top}DZ$. Also, we have $W = \text{diag}\{w_1,\ldots,w_n\}$, $C = \text{diag}\{c_1,\ldots,c_n\}$ and $D = \text{diag}\{d_1,\ldots,d_n\}$, where
\begin{align*}
w_t = & \frac{(1-\sigma_t^2)}{\sigma_t^2}\left[\psi'\!\!\left(\frac{\mu_t(1-\sigma_t^2)}{\sigma_t^2}\right) + \psi'\!\!\left(\frac{(1-\mu_t)(1-\sigma_t^2)}{\sigma_t^2}\right)\right]\frac{1}{\left[g'(\mu_t)\right]^2}, \\
c_t = & \frac{(2-2\sigma_t^2)}{\sigma_t^5}\left[\mu_t\psi'\!\!\left(\frac{\mu_t(1-\sigma_t^2)}{\sigma_t^2}\right) -(1-\mu_t) \psi'\!\!\left(\frac{(1-\mu_t)(1-\sigma_t^2)}{\sigma_t^2}\right)\right], \\
d_t = & \frac{4}{\sigma_t^6}\!\left[\!\mu_t^2\psi'\!\!\left(\!\frac{\mu_t(1\!-\!\sigma_t^2)}{\sigma_t^2}\!\right)\!\! -(1\!-\!\mu_t)^2 \psi'\!\!\left(\!\frac{(1\!-\!\mu_t)(1\!-\!\sigma_t^2)}{\sigma_t^2}\!\right)
\!\!-\psi'\!\!\left(\!\frac{(1\!-\!\sigma_t^2)}{\sigma_t^2}\!\right)\!\right] \frac{1}{\left[h'(\sigma_t)\right]^2}. 
\end{align*}
\normalsize





\bibliographystyle{elsarticle-harv}
\bibliography{bootstrap_AIC}







\end{document}